\pdfoutput=1
\documentclass[journal = jctcce]{achemso}
\usepackage{amsmath}
\usepackage{amssymb}
\usepackage{dcolumn} 
\usepackage{setspace}
\usepackage{array}
\usepackage{graphicx}
\usepackage{booktabs}
\usepackage{gensymb}
\usepackage{multirow}
\usepackage{psfrag}
\nocite{*} 
\newcommand{\eqn}[1]{Eq.\ (\ref{#1})}

\def\br{{\mathbf{r}}}

\def\bS{{\mathbf{S}}}

\def\rhor{{\rho({\bf r})}}

\def\rhoi{{\rho_I}}

\def\rhoir{{\rho_I({\bf r})}}

\author{Pablo Ramos}
\author{Markos Papadakis}
\author{Michele Pavanello}
\email{m.pavanello@rutgers.edu}
\affiliation{Department of Chemistry, Rutgers University, Newark, NJ 07102, USA}
\title{Performance of Frozen Density Embedding for Modeling Hole Transfer Reactions}

\begin{document}
\maketitle
\section*{Abstract}
We have carried out a thorough benchmark of the FDE-ET method for calculating hole transfer couplings. We have considered 10 exchange-correlation functionals, 3 non-additive kinetic energy functionals and 3 basis sets. Overall, we conclude that with a 7\% mean relative unsigned error, the PBE functional coupled with the PW91k non-additive Kinetic energy functional and a TZP basis set constitutes the most stable, and accurate level of theory for hole-transfer coupling calculations. The FDE-ET method is found to be an excellent tool for computing diabatic couplings for hole transfer reactions.

\newpage
\section{Introduction}
The quantum mechanical study of realistically sized molecular systems has become a goal for quantum chemistry and material science. To that aim, multilevel and multiscale algorithms (such as QM/MM) generally approach the problem by representing the system as a set of subsystems whose interaction is accounted for approximately. Along these lines, the frozen density-embedding (FDE, hereafter) formalism developed by Wesolowski and Warshel\cite{weso1993,weso2006} (see Ref.\ \citenum{jaco2013} for a recent review), has become a popular avenue of research. FDE has been applied to a vast array of chemical problems, for instance, solvent effects on different types of spectroscopy\cite{neug2005b,neug2010a,HD2014}, magnetic properties \cite{jaco2006b,bulo2008,neug2005f,kevo2013,weso1999}, excited states\cite{neug2010,casi2004,pava2013b,neug2005b,garc2006},  charge transfer states \cite{ramos2014,pava2013a,solo2014}. Computationally, FDE is available for molecular systems  in ADF \cite{adf,jaco2008b}, Dalton \cite{DALTON,hofn2012a}, and Turbomole \cite{Turbomoleb,ahlr1989,lari2010} packages, as well as for molecular periodic systems in CP2K \cite{cp2k,iann2006} and fully periodic systems (although in different flavors) in CASTEP \cite{CASTEP,laha2007}, Quantum Espresso \cite{fdeinqe,qe,Ale2014}, and Abinit \cite{ABINIT,govi1998}. 

The FDE method casts itself in the framework of subsystem density functional theory, by which the electron density of the total system is split into subsystem contributions and can be determined by solving coupled equations featuring an effective embedding potential. In this way, polarization given by the interaction of the subsystems is included.  This subdivision of the total electron density into subsystem contributions has lead to the use of FDE as an effective charge and spin localization technique \cite{pava2013a,solo2014,solo2012,pava2011b}. Although the reasons for the ability of FDE to yield charge localized states will be addressed in the following section, here we will take this for granted and discuss why such a property of this method is of interest.

The quest for computing charge localized electronic states has a long history \cite{london1932a,cave1997,newton1991,Creutz1994,newt1984,wars1980,marcus1985}, especially in recent years with the advent of the Generalized Mulliken--Hush method\cite{cave1997} and constrained DFT \cite{kadu2012,vanv2010a}. Charge localized states are also known as diabatic states because it has been shown that they minimize the corresponding non-adiabatic coupling matrix \cite{subo2009,pava2011a} -- a defining property of diabatic states \cite{london1932a,mead1982}. Modeling of charge transfer (CT) reactions often involves the use of only two diabatic states: a state where the charge is on the donor (D) also called initial state, and a state where the charge is on the acceptor (A), also known as final state. However, because the charge localized states are not necessarily eigenstates of the molecular Hamiltonian of the system, nor they are constructed to be orthogonal, we expect the Hamiltonian and overlap matrix to be non-diagonal. Namely,
\begin{equation}
\label{hands}
 \mathbf{H}=\left(
\begin{array}{cr}
  H_{DD} & H_{DA}\\
  H_{AD} &  H_{AA} 
\end{array}
\right),~~
  \mathbf{S}=\left(
\begin{array}{cr}
  1 &  S_{DA}\\
 S_{AD} &  1
\end{array}
\right).
\end{equation}
Whether the CT rate is computed with Marcus theory \cite{marcus1956} or it is extracted from a non-adiabatic dynamics \cite{nitzan_book,zene1932}, it is related to the following matrix element \cite{efri1976,migliore2011a}:
\begin{equation}
 \label{coulow}
 V_{DA} = \frac {1}{1-S_{DA}^2} \left(H_{DA}-S_{DA}\frac{H_{DD}+H_{AA}}{2}\right),
\end{equation}
which is known as {\it transfer integral}, {\it coupling matrix element},  {\it charge transfer coupling}, etc.  For many systems of interest,  $V_{DA}$ depends strongly on the molecular geometry \cite{voit2008b,beratan2014,skou2010}, thus the dynamic charge transfer process is more straightforwardly modeled with real-time dynamics methods, such as Tully's surface hopping or Ehrenfest dynamics  \cite{tully1998} as advocated by many \cite{kuba2013,voityuk2012}. For such a model to be computationally efficient, the electronic couplings between the diabatic states have to be computed efficiently. Several methods have been proposed for this task, such as semiempirical methods \cite{fruz2014,guti2009,Priya1996,skou1993}, methods exploting the frozen orbital approximation \cite{groz2000,poel2013} (i.e., the molecular orbitals of the diabatic states are approximated by the ones of isolated donor and acceptor fragments), or all-electron methods such as wavefunction methods which are subsequently rotated to yield diabatic states \cite{cave1997,subo2009,voityuk2012}, or the ones that focus on constructing the diabatic states by imposing locality of the electronic structure \cite{kadu2012,wu2006,pava2011b,pava2013a}.

This explosion of methods for calculating the coupling matrix elements between diabatic states for electron transfer processes called for the setup of a benchmark set which can be used by all researchers developing novel algorithms. In addition, a question which is frequently posed is whether charge transfer couplings are sensitive to the particular method chosen for their evaluation. These questions were posed to an audience at the 2011 conference on ``Charge Transfer in Biosystems'' organized by Profs.\ Rosa di Felice and Marcus Elstner. The seed planted in that conference gave rise to an interesting work by Kubas {\it et al.} \cite{Kubas2014}, where the authors compared benchmark values of coupling elements for 11 molecular dyads calculated with correlated wavefunction methods associated with the Generalized Mulliken--Hush diabatization with values computed with the constrained density functional (CDFT) method, fragment-orbital DFT (FODFT) and density functional tight-binding (FODFTB). The test set was named HAB11, and it was found that the all-electron CDFT method (as implemented by Oberhofer {\it et al.}\cite{ober2010} in the CPMD software \cite{marx_book,andreoni_cpmd}) can reach a mean relative unsigned error (MRUE) of 5\% if it is employed in conjunction with 50\% of Hartree--Fock (HF) exchange in the PBE functional, and a deviation of about 39\% if the pure PBE functional was used. The non-selfconsistent fragment orbital method yielded a deviation of 38\% and the semiempirical FODFTB of 42\%. 

HAB11 consists of eleven $\pi$-conjugated dimers, plus four additional aromatic rings. In Table \ref{alldimers} the structures of the monomers are shown. These organic molecules were chosen because they feature different $\pi$ bond arrangements and different kinds of heteroatoms. The monomers are: ethylene, acetylene, cyclopropene (having one high electronic density bond), the antiaromatic ring cyclobutadiene, O, N and S containing heterocycles, five polycyclic aromatic hydrocarbons (benzene, naphthalene, anthracene, tetracene and pentacene), and one derivative of benzene: phenol. These organic compounds are well known to be part of efficient semiconductor materials\cite{goddar2004,hualong2008,Karsten2011,Troisi2011} and some take part in CT processes in biomolecules\cite{Hunter2001,grey1996}. We obtained the Cartesian coordinates for every structure from the reference\cite{Kubas2014} (for details about how the geometries were obtained we refer the reader to that source). 

The purpose of this work is to provide the community with information about the performance of the FDE method against the HAB11 benchmark set. As the FDE method is used to obtain the electronic structure of the diabats, a post-SCF calculation follows for the determination of the couplings \cite{pava2011b,pava2013a,solo2014}. The resulting composite method is termed FDE-ET, hereafter\cite{solo2014}.  As we will see in the results section, the FDE-ET method compares quite well with the benchmark calculations, albeit experiencing outliers, in most part traceable back to convergence issues. To offer as complete of a picture as we can, the FDE coupling calculations are carried out with 10 different exchange correlation density functional(XC, hereafter), 3 non-additive Kinetic energy functionals (NAKE, hereafter) and 3 basis sets. This totals to a staggering 90 levels of theory tested in this work, leading to a total of 5400 coupling calculations for the HAB11 set alone. In addition, and following Ref.\ \citenum{Kubas2014}, we have carried out calculations for dyads whose monomers were rotated with respect to each other, presenting an additional 3780 coupling calculations.

This paper is organized as follows, in section \ref{FDEt} we show briefly the characteristics of FDE-ET. In section \ref{CD} the computational details are described. Section \ref{results} collects the results of the comparisons against the HAB11 test set and for rotated ethylene and thiophene dimers. Finally, in section \ref{conclu} we outline the conclusions. 

\begin{table}[htp]
\begin{center}
\begin{tabular}{cccc}
\toprule
Dimer & Symbol$^a$ & Structure & Reference method$^b$\\
\hline
Ethylene  & (EE) & \includegraphics[width=0.1\textwidth]{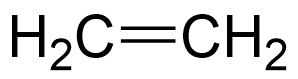} & MRCI+Q\\
Acetylene  & (AC) & \includegraphics[width=0.1\textwidth]{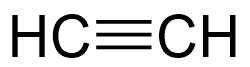} &  MRCI+Q\\
Cyclopropene  & (CP) & \includegraphics[width=0.05\textwidth]{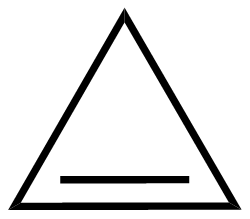} & MRCI+Q\\
Cyclbutadiene  & (CB) & \includegraphics[width=0.05\textwidth]{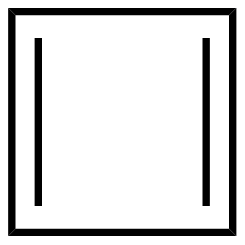} & MRCI+Q\\
Cyclopentadiene  & (CD) & \includegraphics[width=0.05\textwidth]{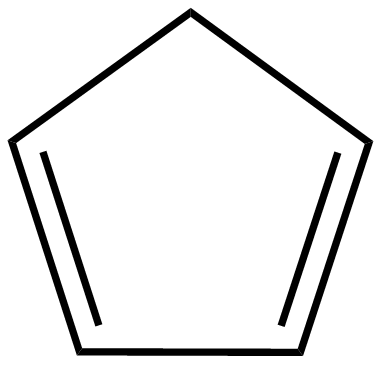} & MRCI+Q\\
Furane  & (FF) & \includegraphics[width=0.05\textwidth]{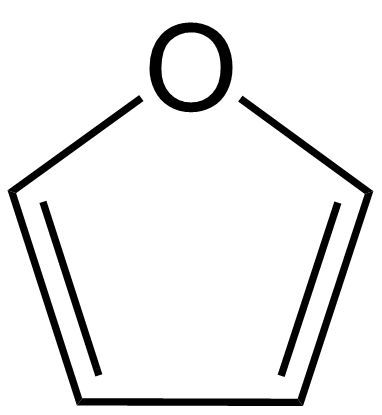} & MRCI+Q\\
Pyrrole  & (PY) & \includegraphics[width=0.05\textwidth]{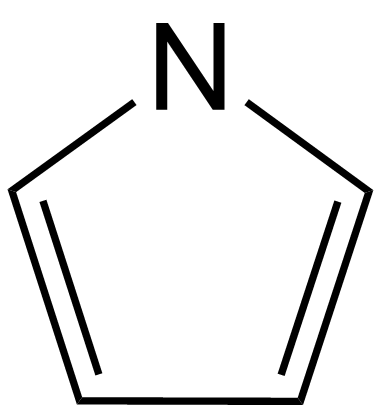} & MRCI+Q\\
Thiophene  & (TH) & \includegraphics[width=0.05\textwidth]{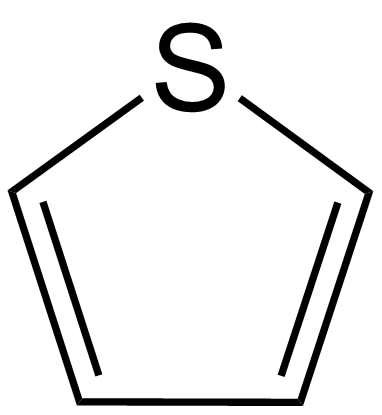} & NEVPT2 \\
Imidazole  & (IM) & \includegraphics[width=0.05\textwidth]{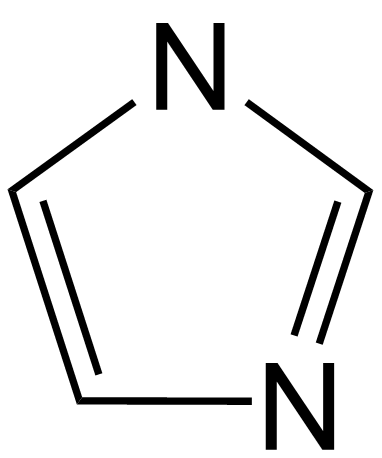} & NEVPT2 \\
Phenol &  (PH) & \includegraphics[width=0.05\textwidth]{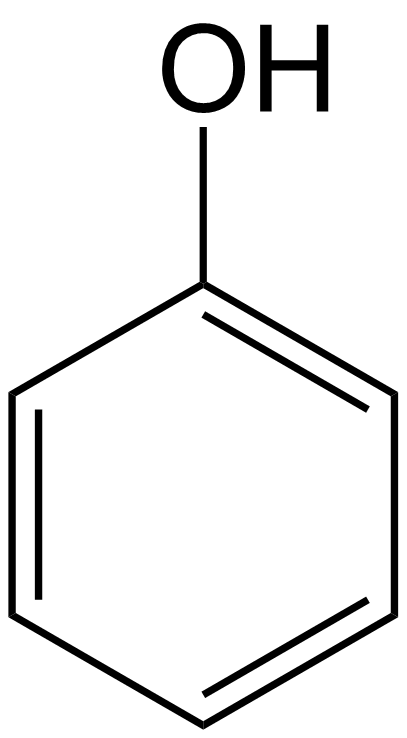} & NEVPT2 \\
Benzene  & (BB) & \includegraphics[width=0.05\textwidth]{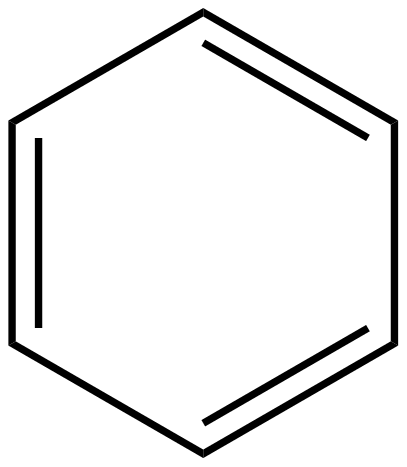} & NEVPT2 \\
Naphthalene &  (NN) & \includegraphics[width=0.1\textwidth]{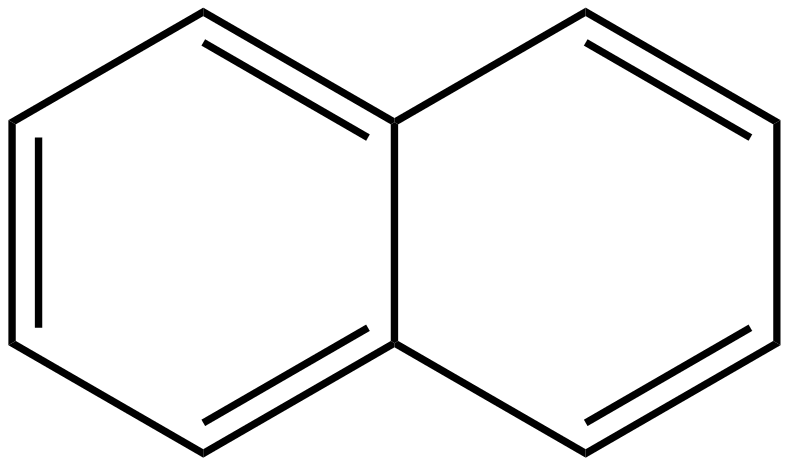} & SCS-CC2\\
Anthracene &  (AA)& \includegraphics[width=0.15\textwidth]{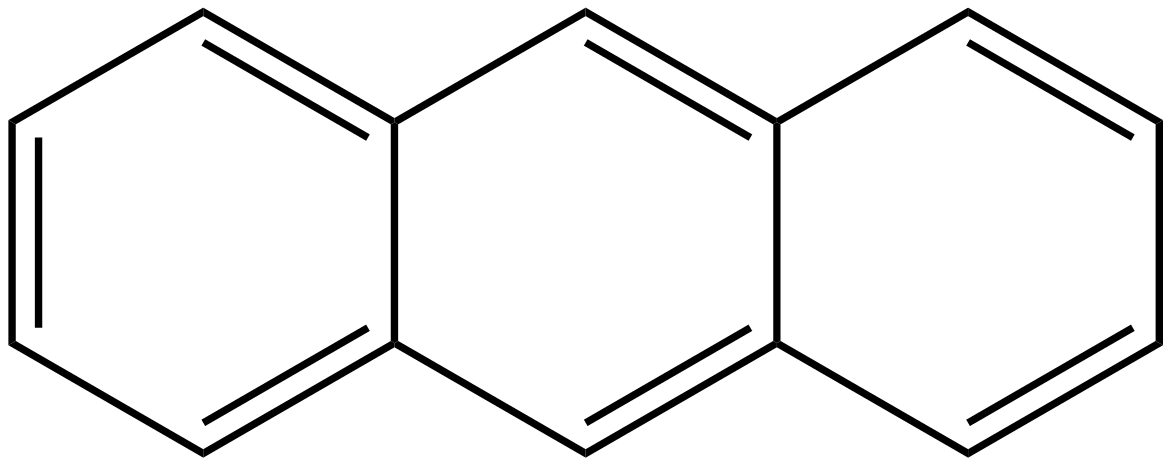} & SCS-CC2\\
Tetracene &  (TT) & \includegraphics[width=0.2\textwidth]{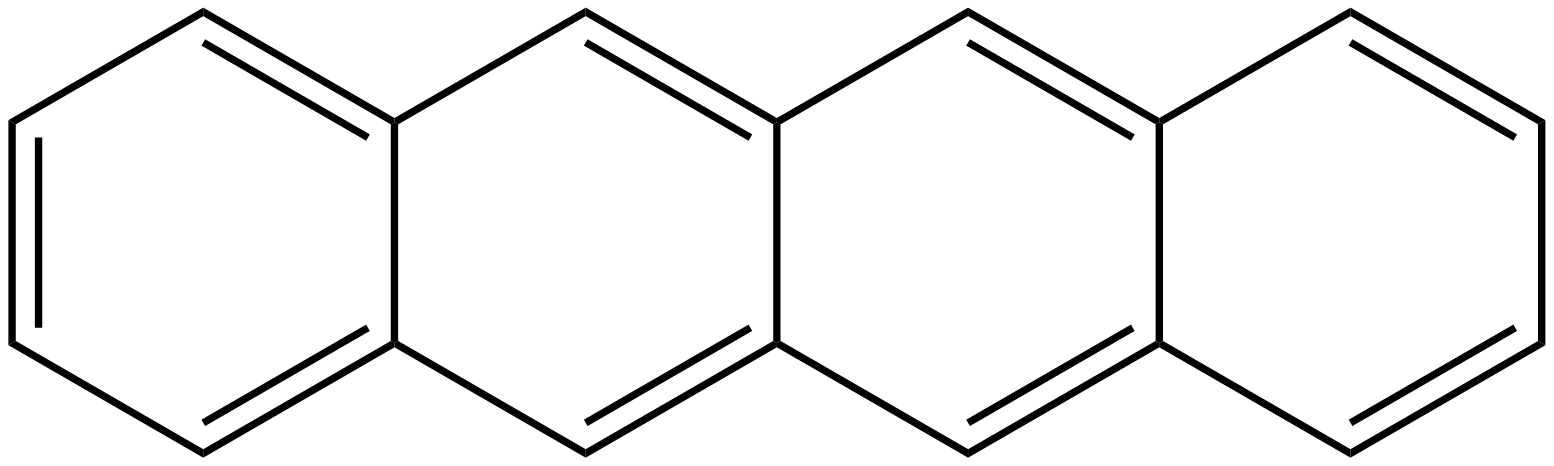} & SCS-CC2\\
Pentacene  & (PP) & \includegraphics[width=0.25\textwidth]{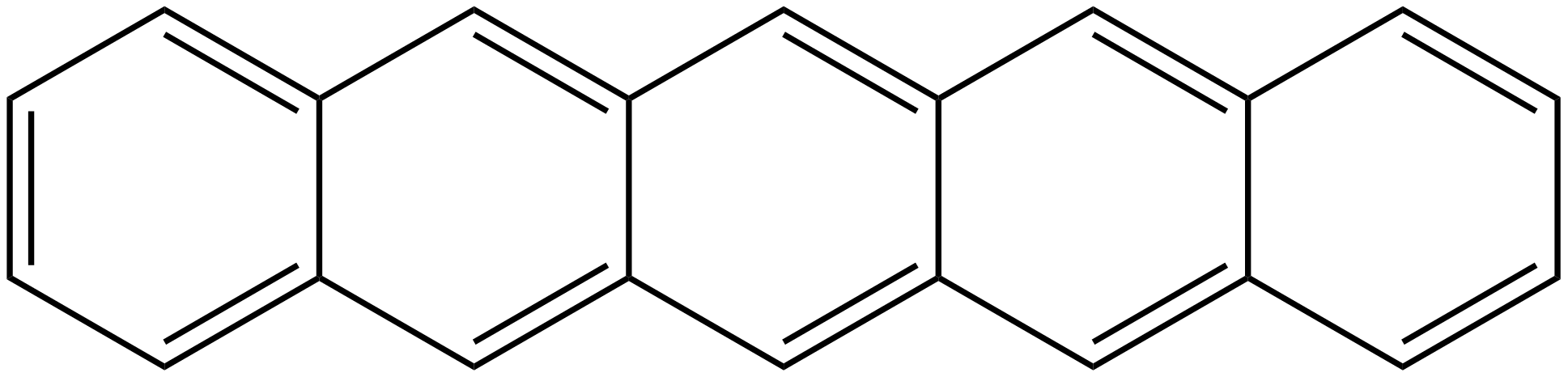} & SCS-CC2\\
\bottomrule
\end{tabular}
\end{center}
\begin{flushleft}
\begin{center}
 $a$ Abbreviations used in this work.\\
 $b$ Ref.\citenum{Kubas2014} 
\end{center}
\end{flushleft}
\caption{\label{alldimers} Dimers of the HAB11 test set.}
\end{table}

\section{Diabatic states from Frozen Density Embedding}
\label{FDEt}
\subsection{Background on FDE}
The FDE formalism prescribes the total electron density to be expressed as the sum of subsystem electron densities\cite{weso1993,sen1986,cort1991}. Namely, 
\begin{equation}
 \label{fdeden}
 \rho_{tot}(r)=\sum_{I=1}^{N_s} \rho_I(r).
\end{equation}

Where $N_s$ is the number of subsystems.

The electron density of each subsystem is obtained by solving a Kohn--Sham (KS) like equation augmented by an embedding potential that accounts for the interactions of other subsystems whose density is kept frozen in this step, such as
\begin{equation}
 \label{ksequ}
 \left[\frac{-\nabla^2}{2}+\upsilon^{I}_{KS}(r)+\upsilon^{I}_{emb}(r)\right]\phi_{(i)I}(r)=\epsilon_{(i)I}(r)\phi_{(i)I}(r).
\end{equation}
Where $\phi_{(i)I}(r)$ are the molecular orbitals of subsystem $I$, and $\upsilon^{I}_{emb}(r)$ is the embedding potential acting on the same subsystem reading as follows:
\begin{align}
 \label{embpot}
 \nonumber
 \upsilon^{I}_{emb}(r)=&\sum^{N_s}_{J\neq I}\left[\int \frac{\rho_J}{|r-r'|}dr'-\sum_{\alpha\in J} \frac{Z_{\alpha}}{|r-R_{\alpha}|}\right]+ \\
&+\frac{\delta T_{\rm s}[\rho]}{\delta\rhor}-\frac{\delta T_{\rm s}[\rhoi]}{\delta\rhoir}+\frac{\delta E_{\rm xc}[\rho]}{\delta\rhor}-\frac{\delta E_{\rm xc}[\rhoi]}{\delta\rhoir}.
\end{align}
In the above, $T_{\rm s}$, $E_{\rm xc}$ and $Z_\alpha$ are kinetic and exchange--correlation energy functionals, and the nuclear charge, respectively. A special comment for the kinetic energy is needed. In the KS method, $T_{\rm s}[\rho]$ should be calculated from the molecular orbitals of the entire system. However, these orbitals are not calculated in FDE and therefore are not accessible. Approximate kinetic energy functionals are employed instead, representing the NAKE term with a semilocal functional. This approximation is ultimately the biggest difference between an FDE and a full KS-DFT calculation of the supersystem\cite{gotz2009,weso1996,Lude2002}. For example, when the subsystems feature a large overlap between their electron densities, FDE in conjunction with GGA NAKE functionals becomes inaccurate when compared to regular KS-DFT \cite{Ale2014,fux2008,kiew2008}. To achieve selfconsistency, the subsystem densities are determined in an iterative way called freeze-and-thaw \cite{weso1996b,jaco2008b}. 

\subsection{How does FDE generate diabats?}
Diabatic states can be generated with FDE by construction. In practical terms, the calculation is performed on at least two interacting subsystems (donor and acceptor fragments) whose electron densities are determined via the freeze-and-thaw procedure employing approximated NAKE functionals. A set of two simulations is set up: one featuring a hole on the donor fragment (i.e., the KS-like equations are solved imposing the density of the corresponding subsystem to integrate to a number of electrons defecting by one compared to the neutral fragment), and another calculation in which the acceptor is now positively charged. We should note that it is also possible to increase the number of electrons by one -- in this case, an excess electron is generated on the subsystem rather than a hole. The result of the calculations is that the hole (or electron) is completely localized onto the fragment it was placed on at the onset of the calculation. 

\begin{figure}
\includegraphics[width=0.7\textwidth]{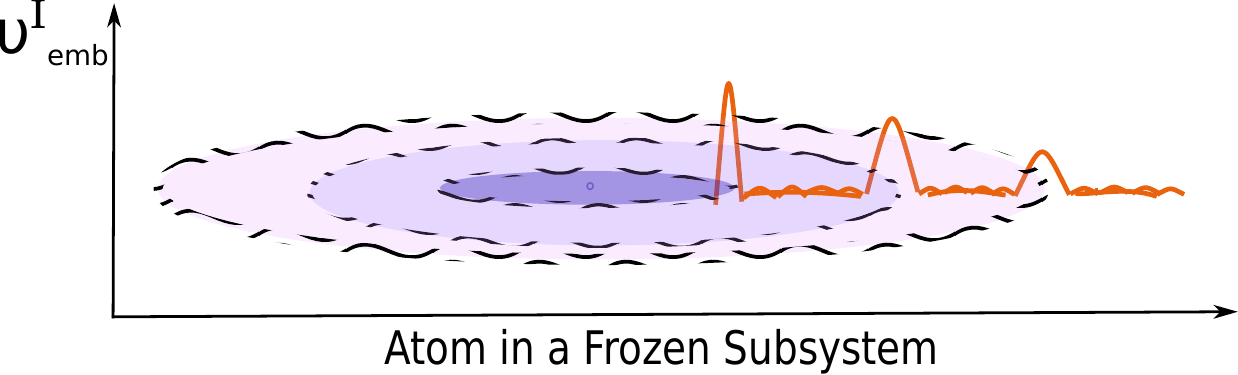}
\caption{\label{nake}Depiction of the shape of the embedding potential (in red) in the region of the atomic shells of surrounding subsystems (aka frozen subsystems).}
\end{figure}

There are four reasons for the FDE calculations to yield charge localize states \cite{pava2013a}. First, the subsystem orbitals are not imposed to be orthogonal to orbitals of the other subsystems. This is important as it implies that not imposing orthogonality removes a bias towards delocalization, as noted by Dulak and Wesolowski \cite{dula2006}. However, this reason alone is not enough. A second reason is the fact that FDE calculations are carried out in the monomer basis set [i.e., using the FDE(m) method \cite{jaco2007}]. With no basis functions on the surrounding frozen subsystems, a charge transfer between the subsystems becomes an unlikely event and the SCF is biased to converge to a charge localized solution. The third reason is similar to the previous one and invokes the fact that FDE calculations are always initiated with a subsystem localized guess density. The initial conditions also have a bias effect on the final SCF solution -- a localized initial guess density will likely yield an SCF solution that is subsystem localized as well. The fourth reason is more subtle. It deals with the shape of the embedding potential in the region of the surrounding fragments. 
Electrons remain localized also because there are repulsive walls in the vicinity of the atomic shells of atoms belonging to the surrounding subsystems. As noted by Jacob {\it et al.}  \cite{jaco2007}, the approximate kinetic energy functionals are unable to cancel out the attractive potential due to the nuclear charge in the vicinity of the nucleus. However, the shape of most semilocal kinetic energy potentials is such that in going towards the nucleus they start out too low compared to the exact potentials, then cross the exact potential and become larger in the region of an atomic shell. This is so up to when the shell has faded, then the potential becomes again too attractive (see for example Figures 4 and 5 of Ref.\ \citenum{wang2000} and Figure \ref{nake} for a simplified depiction of this effect). This behavior was also reported by Fux {\it et al.}\cite{fux2010} when they calculated the approximate vs.\ accurate potentials for selected dyads. 

Until now, we have assumed that the the SCF procedure in FDE (i.e.\ when the so-called freeze-and-thaw cycles are used, see the computational detals) leads to a unique solution. This has been challenged recently in the limit of exact NAKE functionals \cite{grit2013}. In this work, however, we only employ approximate functionals for which experience shows that a unique solution is always found. Numerical evidence of this can be found in Ref.\cite{nafz2014}.
\subsection{Coupling calculations with FDE: the FDE-ET method}
In case of non-orthogonal functions, the coupling (exact at the Hartree--Fock level but only approximate at the DFT level) can be expressed as \cite{pava2013a,thom2009a,Evan2013}:
\begin{equation}
\label{cou}
 H_{DA}=\langle \psi_{D} | \hat H |\psi_{A}\rangle=S_{DA}E\left[\rho^{(DA)}(\br)\right].
\end{equation}
where $\hat H$ is the molecular electronic Hamiltonian, $\psi_{D}$ \ and $\psi_{A}$ are the two diabatic states (D for donor, A for acceptor) and  $\rho^{(DA)}(\br)$ is the transition density which reads as:
\begin{equation}
\label{tranrho}
 \rho^{(DA)}(\br)=\langle \psi_D | \sum_{k=1}^{\# of electrons} \delta (\br_k-\br) | \psi_A\rangle
\end{equation}
Assuming the wavefunctions to be expressed in terms of single Slater determinants, the overlap element appearing in \eqn{tranrho} is found by computing the following determinant:
\begin{equation}
\label{eq:s}
 S_{DA}=\mathrm{det}\left[\mathbf{S}^{(DA)}\right],
\end{equation}
where $\bS^{DA}_{kl}=\langle \phi_{k}^{(D)} | \phi_{l}^{(A)} \rangle$ is the transition overlap matrix in terms of the occupied orbitals ($\phi_{k/l}^{(D/A)}$)\cite{maye2002,thom2009a}. Thus, the transition density is now written in the basis of all occupied orbitals which make up the diabatic states  $\psi_{D}$ \ and $\psi_{A}$. Namely,
\begin{equation}
\label{Ivth}
 \rho^{(DA)} (\br)= \sum_{kl}^{\rm occ} \phi_{k}^{(D)} (\br)\left(\mathbf{S}^{(DA)}\right)_{kl}^{-1} \phi_{l}^{(A)}(\br).
\end{equation}
Finally, the Hamiltonian coupling is calculated by plugging \eqn{Ivth} and \eqn{eq:s} in \eqn{cou} and the resulting matrix elements in \eqn{coulow} -- that is, the coupling of two L\"{o}wdin orthogonalized $\psi_D$ and $\psi_A$\cite{lowd1963}. 
In the FDE-ET method, the $\phi_{k}^{(D)}$ and $\phi_{k}^{(A)}$ orbitals are borrowed from the FDE subsystem orbitals as prescribed by Refs.\ \cite{pava2011b,pava2013a}.

\section{Computational Details}
\label{CD}
All effective couplings were calculated with the Amsterdam Density Functional (ADF) package\cite{teve2001a} (2014 release). For this benchmark study, we selected ten XC functionals as follows: three GGAs (PBE\cite{PBEc}, BLYP\cite{beck88,lee1988} and PW91\cite{perd1991}), two MetaGGAs (M06-L\cite{zhao2006} and TPSS\cite{Scuseria2003}), three Hybrid functionals (B3LYP\cite{step1994}, BHandH\cite{becke93a} and PBE0\cite{admo1999}) and two MetaHybrid functionals (M06-2X\cite{zhao2008} and M06-HF\cite{zhao2008}). 
These functionals were employed in the FDE calculations. However, the non-additive term for the non-additive exchange--correlation energy functional ($XC^{NADD}$) needed for the embedding potential and functional is computed always at the local or semilocal level\cite{jacob2013} in order to avoid costly OEP type procedures. Specifically, when B3LYP and BHandH were used, we chose BLYP for the $XC^{NADD}$; PBE0 was replaced by PBE and for both MetaGGA and MetaHybrids, PW91 was chosen.      

Three NAKE functionals were used, the LDA Thomas--Fermi\cite{cort1991} functional (TF, hereafter), the GGA functional PW91k\cite{lemb1994} and the gradient expansion approximation (GEA) functional P92\cite{perd1992}. 

Regarding basis sets, we use three Slater-type basis sets, TZP, TZ2P and QZ4P. TZP and TZ2P are double $\zeta$ in the core and triple $\zeta$ in the valence whereas QZ4P is triple $\zeta$ in core and quadrupole $\zeta$ at the valence. Additionally these basis sets are augmented with one, two and four polarization functions respectively\cite{Lenthe2003}. 

The ADF default settings for the self-consistent field (SCF) cycles procedure were used. Also, the Becke\cite{fran2013} numerical integration grid and the ZlmFit\cite{fran2014} density fitting options were set for both FDE and FDE-ET (through the ElectronTransfer keyword) calculations. 

The FDE-ET electronic couplings are obtained first by running an FDE calculation [i.e., solving for \eqn{ksequ}], where the density is minimized by three freeze and thaw cycles, the first one of these cycles was performed using Thomas--Fermi NAKE and the next two were carried out with the corresponding NAKE. Secondly, a post-SCF evaluation of \eqn{coulow} where each term is given by the 2 $\times$ 2 Hamiltonian and overlap matrices of \eqn{hands}. In the FDE-ET post-SCF step, hybrids, metaGGAs, and metahybrid are not yet supported. Hence, the XC functional was changed in the evaluation of \eqn{cou} by a GGA functional. Specifically: for B3LYP and BHandH the BLYP functional was used. The PBE functional was used when PBE0, MetaGGAs and MetaHybrids were employed.

All dimer structures were taken from the HAB11 databases in Kubas {\it et al.} \cite{Kubas2014}. In total, 15 $\pi$-systems perfectly stacked were analyzed, distance dependence calculations of the electronic coupling at 3.50, 4.00, 4.50 and 5.00 \AA\ were performed. In addition, we took ethylene (EE) dimer and thiophene (TH) dimer to compute the dependence of the coupling for different rotations with respect to the center of mass of each monomer in the same way as Kubas {\it et al.}\cite{Kubas2014}.

\section{Results and Discussion}
\label{results}
The FDE-ET couplings are compared for each system against correlated ab-initio wavefunction methods obtained by Kubas {\it et al.} \cite{Kubas2014}. In that work, MRCI+Q, NEVPT2 and SCS-CC2 were employed depending on the system size. MRCI+Q was used for ethylene (EE), acetylene (AC), cyclopropene (CP), cyclobutadiene (CB), cyclopentadiene (CD) and furane (FF) dimers. NEVPT2 was used for pyrrole (PY), thiophene (TH), imidazole (IM), benzene (BB) \footnote[0]{Additional reference data for benzene using SCS-CC2 method is given in reference \citenum{Kubas2014}. In this work we compared with NEVPT2 level of theory.} and phenol (PH); finally, SCS-CC2 was employed for the larger rings, such as naphthalene (NN), anthracene (AA), tetracene (TT), and pentacene (PP). For sake of completeness, in the supplementary information\cite{EPAPS} we report all calculated couplings, all correlation plots and all the error analyses computed for each dimer. There, we report the mean unsigned error (MUE) and mean relative unsigned error (MRUE), the mean relative signed error (MRSE), and the maximum unsigned error (MAX) varying each of the following categories: XC functionals, the NAKE functionals, and the basis sets. To aid our explanation of the results, we have chosen to report in the figures of the main text the variation of the three categories from a common starting point: the PBE/PW91k/TZP level of theory (e.g., XC functional/NAKE/basis set).

\subsection{Effect of the non-additive Kinetic energy functional}
\label{secnake}
\begin{figure}[htp]
\begin{center}
\includegraphics[width=0.8\textwidth]{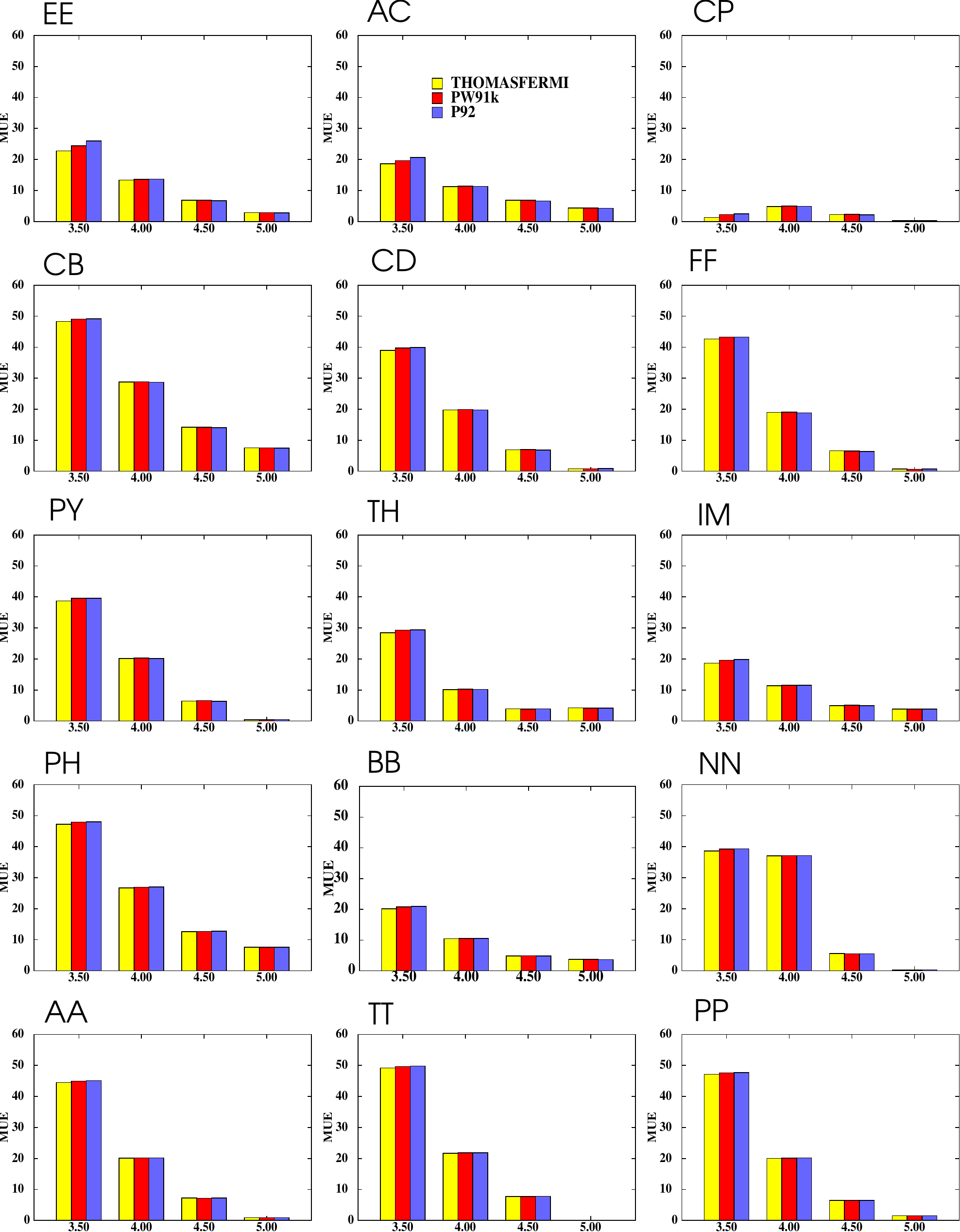} 
\end{center}
\caption{MUE values as a function of different NAKEs used in this study. In this plot, the PBE functional and TZP basis set are employed. Full results are available in the supplementary information section. All bars are in meV.}
\label{muenadd}
\end{figure}
\begin{figure}[htp]
\begin{center}
\includegraphics[width=0.8\textwidth]{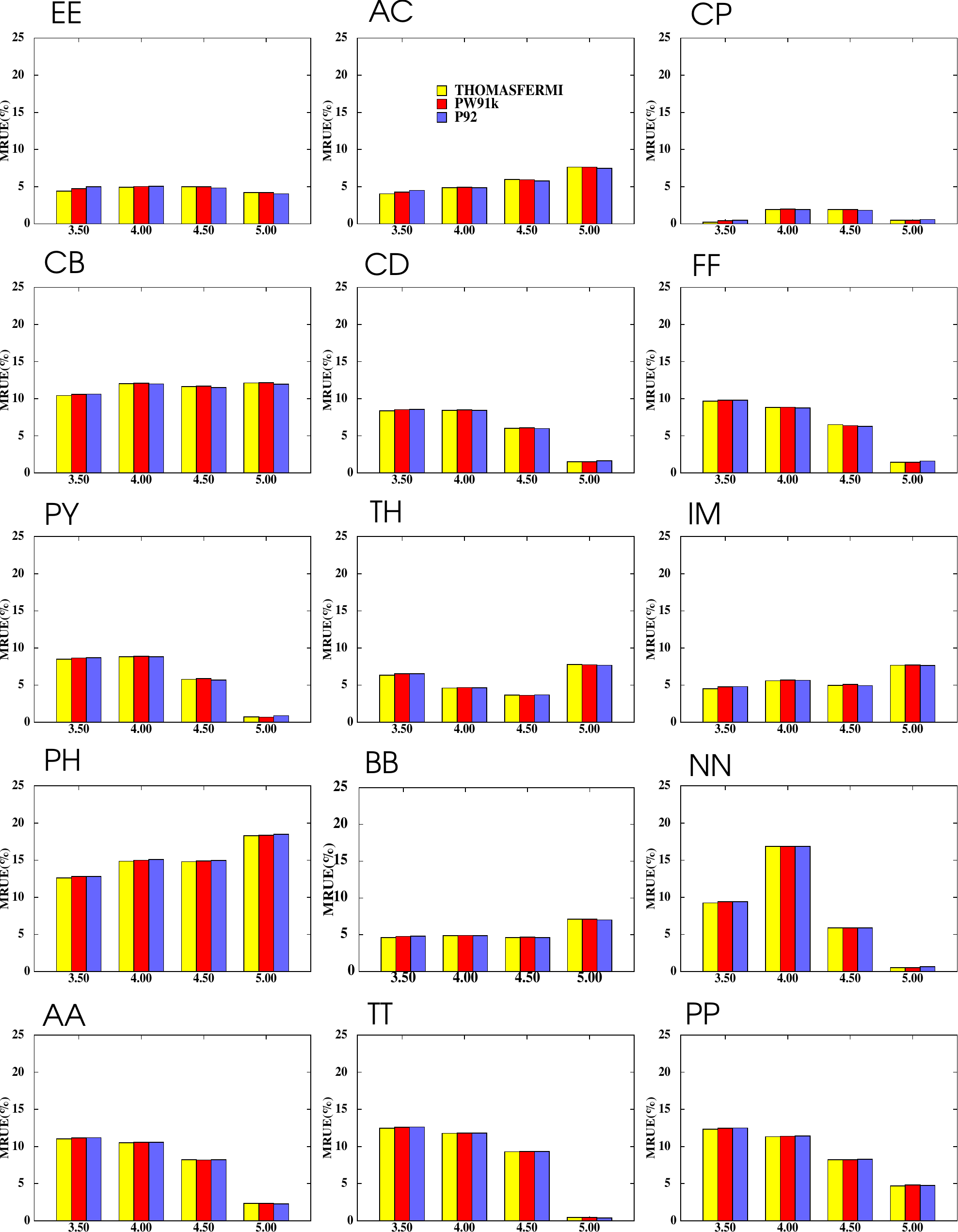} 
\end{center}
\caption{MRUE in the performance of each NAKE functional. The PBE functional and TZP basis set are employed. Full results are available in the supplementary materials section.}
\label{mruenadd}
\end{figure}

Let us first discuss the behavior of the NAKE functional. In Figures \ref{muenadd} (\ref{mruenadd}), the MUE (MRUE) for couplings obtained using the PBE functional and the TZP basis set are reported. All other data are reported in the supplementary materials \cite{EPAPS}. A very good relation with the reference data is clear. The MRUEs are always below 20\% and generally gets better and better as the distance separating the monomers increases. 

Overall, the NAKE functionals are consistent with each other. Each NAKE features couplings that correctly decay exponentially with the inter-monomer distance. For some functionals, we found a not so good description of the electronic coupling for a few dimers (e.g., AC, TT and BB systems in Figure S2 in the supplementary materials), we can see that again the NAKE functionals are consistent and do not change the picture in going from one NAKE to another. 
Nevertheless, in systems like TH or TT the Thomas--Fermi functional performs poorly when used in conjunctions with several XC functionals (Figure S2). We can attribute this behavior to the fact that the Thomas--Fermi potential compared to the GGA NAKE potentials is too soft and is not successful in localizing the hole, especially when the QZ4P basis set is employed. 
\subsection{Effect of the basis set}
\label{secbases}
\begin{figure}[htp]
\begin{center}
\includegraphics[width=0.8\textwidth]{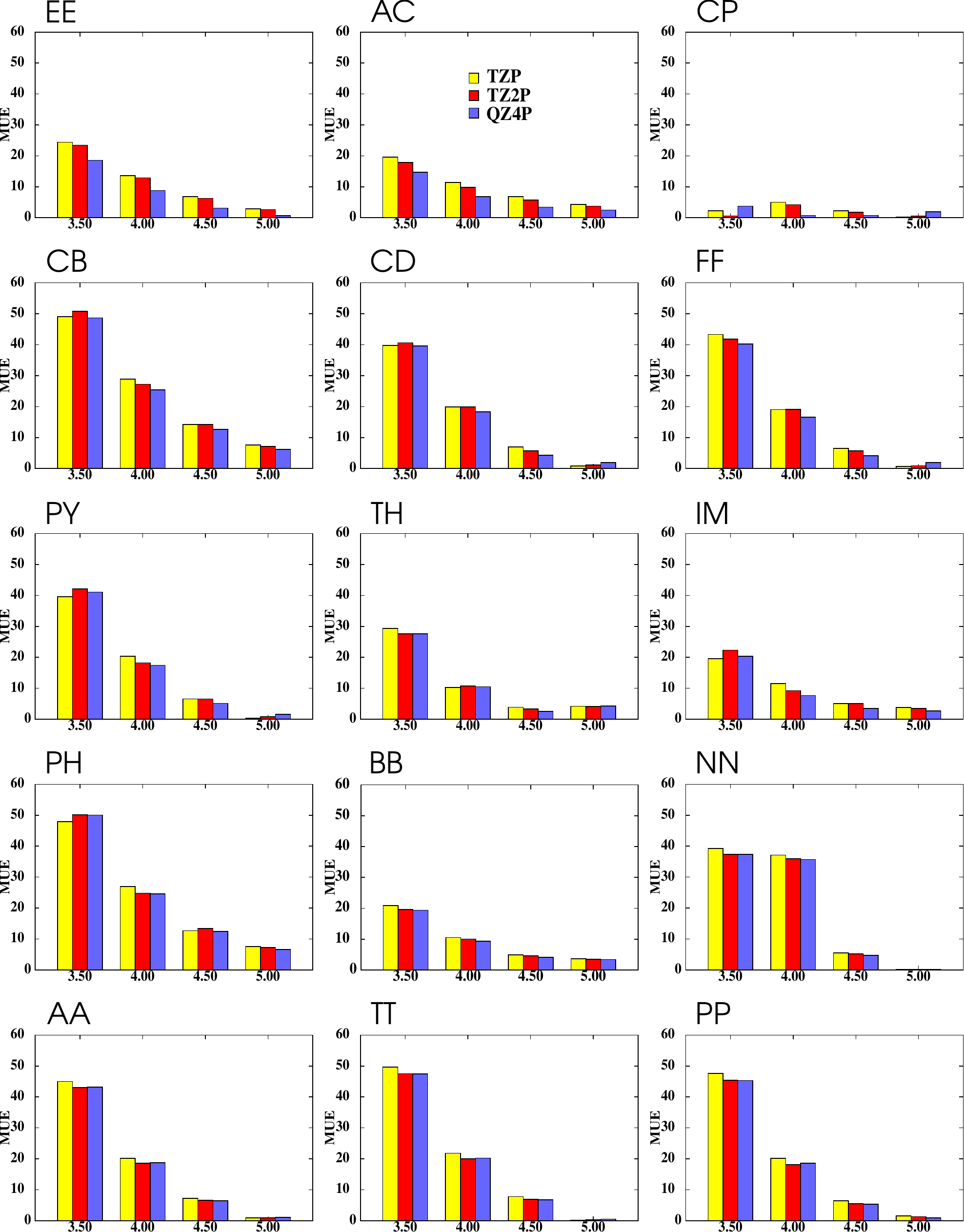} 
\end{center}
\caption{MUE values as a function of the basis sets. PBE and PW91k are employed. See caption to Figure \ref{muenadd} for additional details.}
\label{muebas}
\end{figure}
\begin{figure}[htp]
\begin{center}
\includegraphics[width=0.8\textwidth]{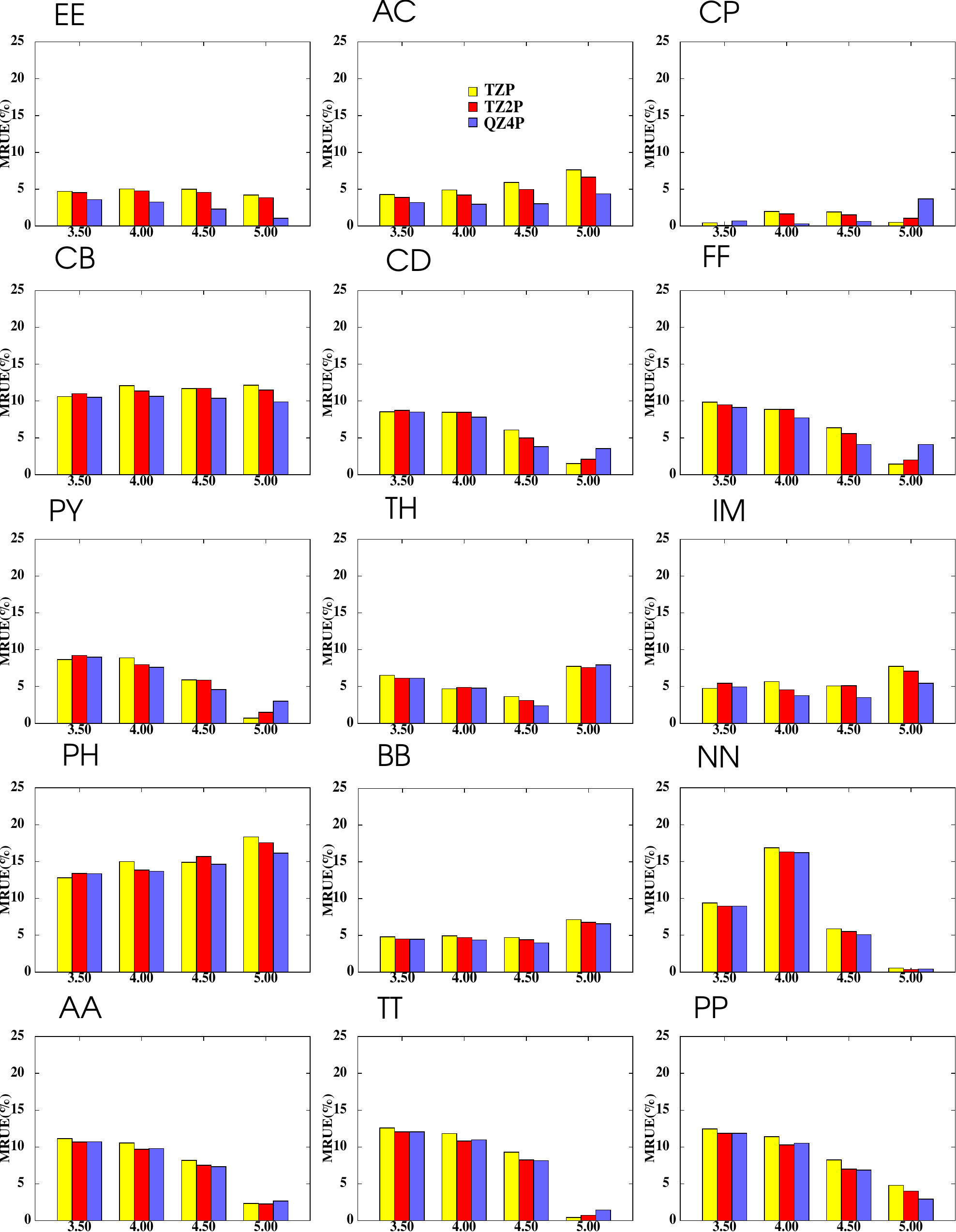} 
\end{center}
\caption{MRUE in \% for the performance of each basis set. PBE and PW91k are employed.}
\label{mruebas}
\end{figure}
We now discuss the effect of the basis set in FDE-ET coupling calculations. Once again, Figures \ref{muebas} (\ref{mruebas}) report the MUE (MRUE) when varying the basis set employing the PBE/PW91k functionals. The picture is not very different from the previous section in the sense that all MRUEs are at or below 20\% with an inclination for being larger for aromatic dimers. Some improvements in the electronic couplings at long distances are noticed when the number of basis functions are increased. This can be explained by the fact that the more diffuse set of functions describes better the tails of the density and allows for a better (closer to the benchmark) coupling calculation. 

In contrast to the choice of NAKE, chosing the basis set has an effect on the accuracy of the calculated couplings. 
When opening the discussion to all the functional/basis set combination considered in this work, deviations are noticeable, especially for the QZ4P basis set for most of the systems at 3.50 \AA. For some dimers/XC functional combinations, there are deviations for all basis sets, see for example benzene dimer (BB) in Figure S3 with the BLYP, PBE0, TPSS, and M06-HF. Thus, we distinguish two scenarios for these deviations, one is when one particular basis fails in conjunction with several XC functionals. We generally see this behavior for the more diffuse QZ4P set and almost never (a few outlier are the exception, such as the metaGGAs for AC and PBE0 for BB) for the other sets. The second situation is when two or all basis sets fail for a certain XC functional. This is attributed to a specific shortcoming of the XC functional. 

In Figure S1 the systems TH, PP, PY, IM and CD show an incorrect coupling at 3.50 \AA\ (where the interaction between the two dyads is the highest) when the QZ4P set is used. Coupling values from 0.1 meV to 1000 meV are reported for these systems (see Table S2 in the supplementary information) while the benchmarks are around 400 meV. The reason for this deviation is that FDE does not impose any constraint to a subsystem calculation to yield a charge localized electronic structure. The four factors responsible for the ability of FDE to yield charge-localized states are found to be systematic in their success of localizing the electronic structure on the subsystems. However, if the monomer basis set is large, one of the four factors becomes less effective, in turn increasing the chances of failure in the localization \cite{solo2014}, or increasing the intersubsystem density overlap, in turn increasing the erroneous behavior of the NAKE \cite{jaco2007}.
\begin{figure}[htp]
\begin{center}
\includegraphics[width=0.8\textwidth]{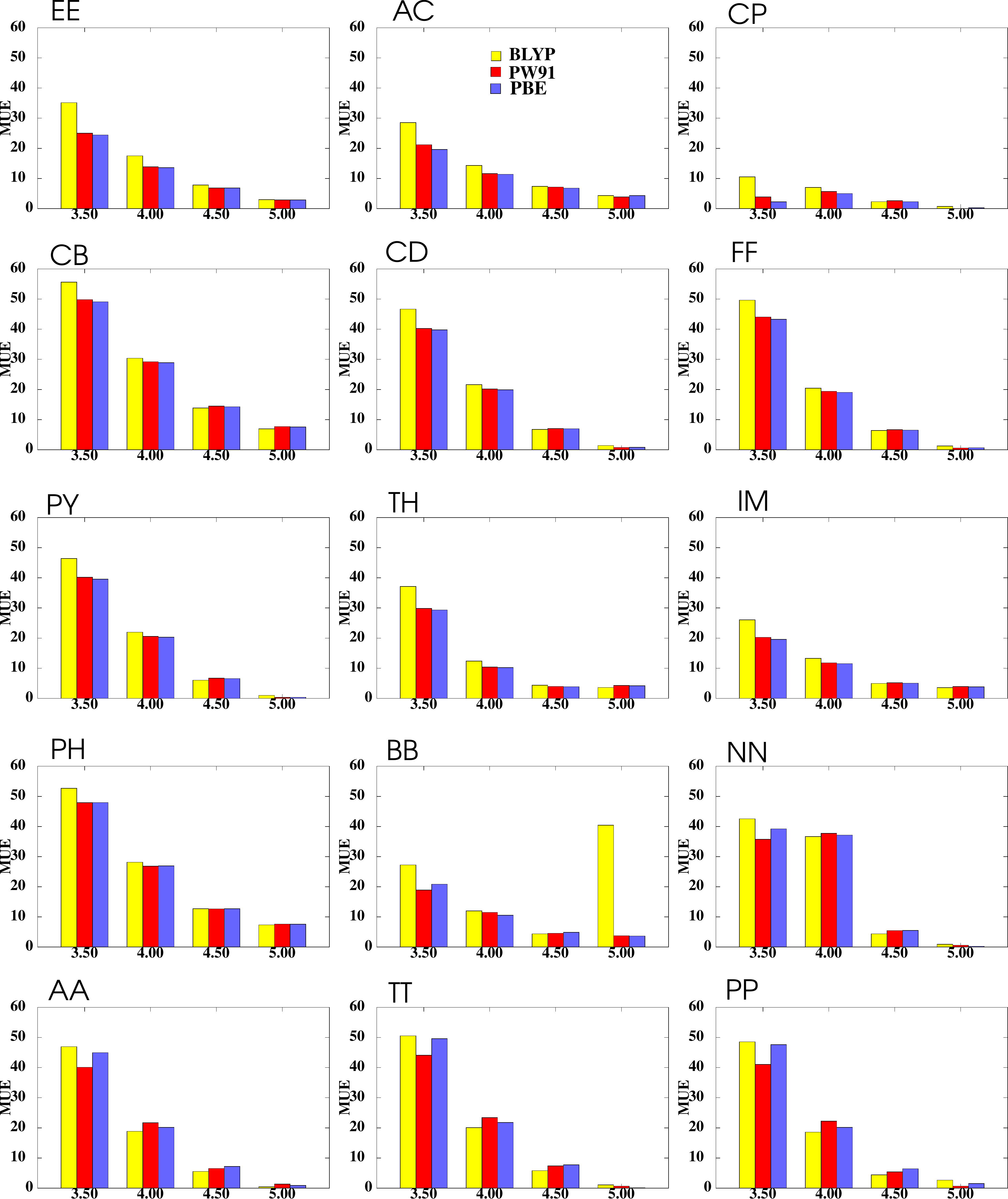} 
\end{center}
\caption{MUE values as a function of GGA XC functionals. TZP and PW91k are employed. See caption to Figure \ref{muenadd} for additional details.}
\label{muefun1}
\end{figure}
\begin{figure}[htp]
\begin{center}
\includegraphics[width=0.8\textwidth]{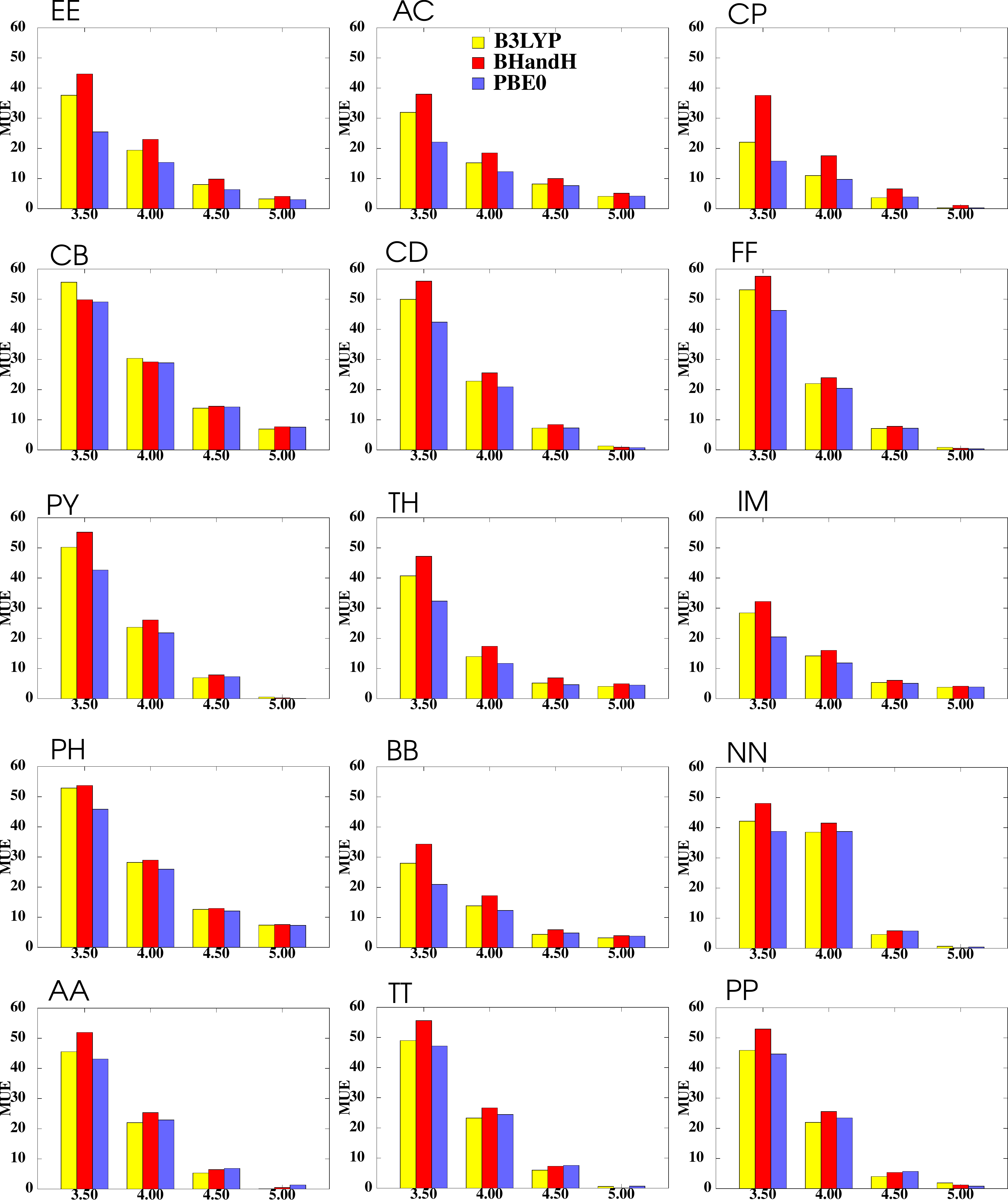} 
\end{center}
\caption{MUE values as a function of hybrid XC functionals. See caption of Figure \ref{muefun1} for additional details.}
\label{muefun2}
\end{figure}
\begin{figure}[htp]
\begin{center}
\includegraphics[width=0.8\textwidth]{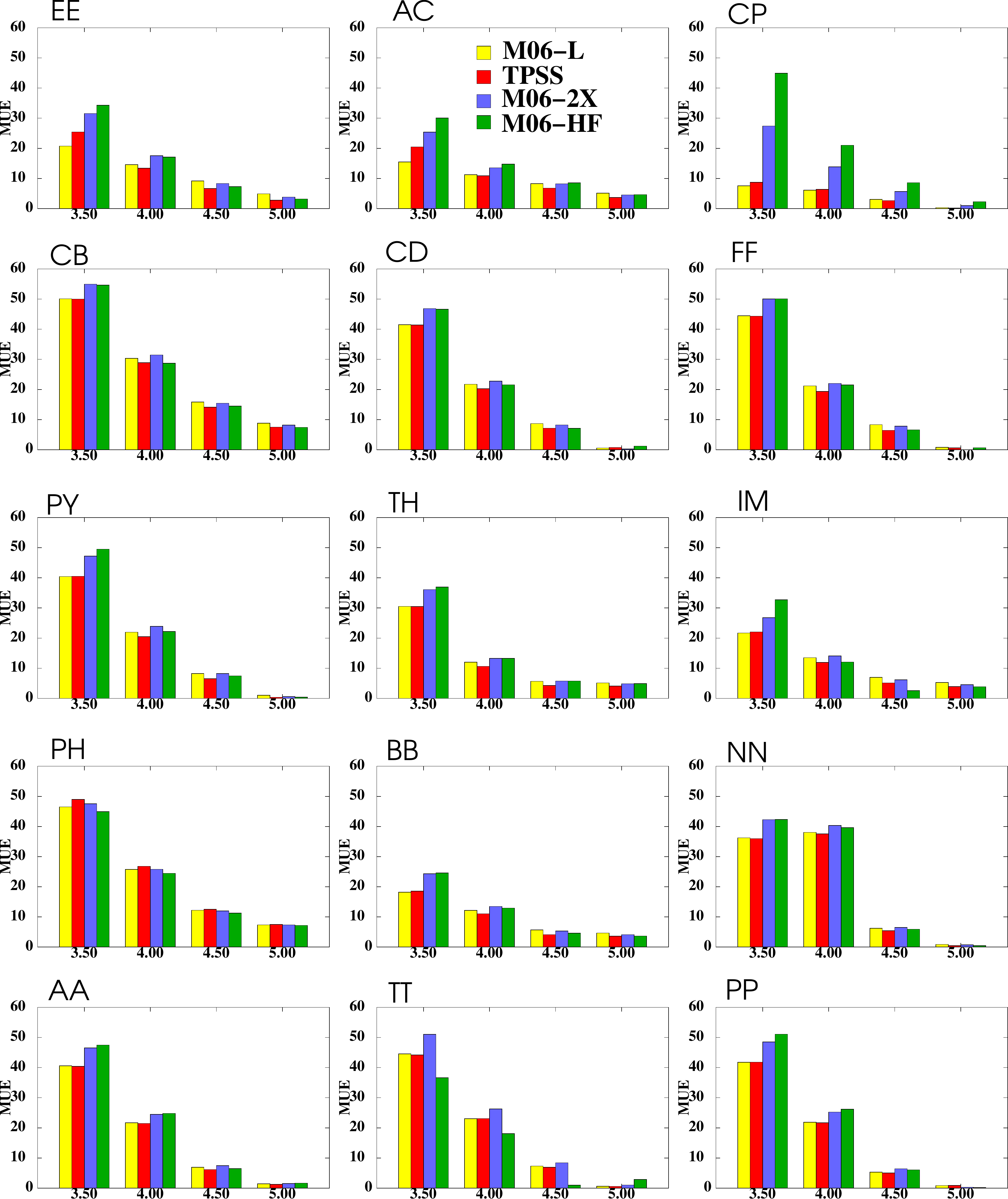} 
\end{center}
\caption{MUE values as a function of the metaGGA and metahybrid functionals. See caption of Figure \ref{muefun1} for additional details.}
\label{muefun3}
\end{figure}

\subsection{Effect of the XC functionals}
\label{statfunc}
This section is devoted to the analysis of the performance of the XC functionals on the calculation of the electronic couplings with the FDE-ET method. Until now the basis set and NAKE functional correlations showed (besides the reported outliers) a relatively insensitive MRUE and MUE distribution along the considered range of distances. As we will see, this is not the case when varying the XC functional. The performance of each functional in all systems is presented in Figures \ref{muefun1} (\ref{mruefun1}), \ref{muefun2} (\ref{mruefun2}), and \ref{muefun3} (\ref{mruefun3}), where MUE (MRUE) is shown. From the figures it is clear that all functionals behave well at the various distances for the majority of the systems.  
\begin{figure}[htp]
\begin{center}
\includegraphics[width=0.6\textwidth]{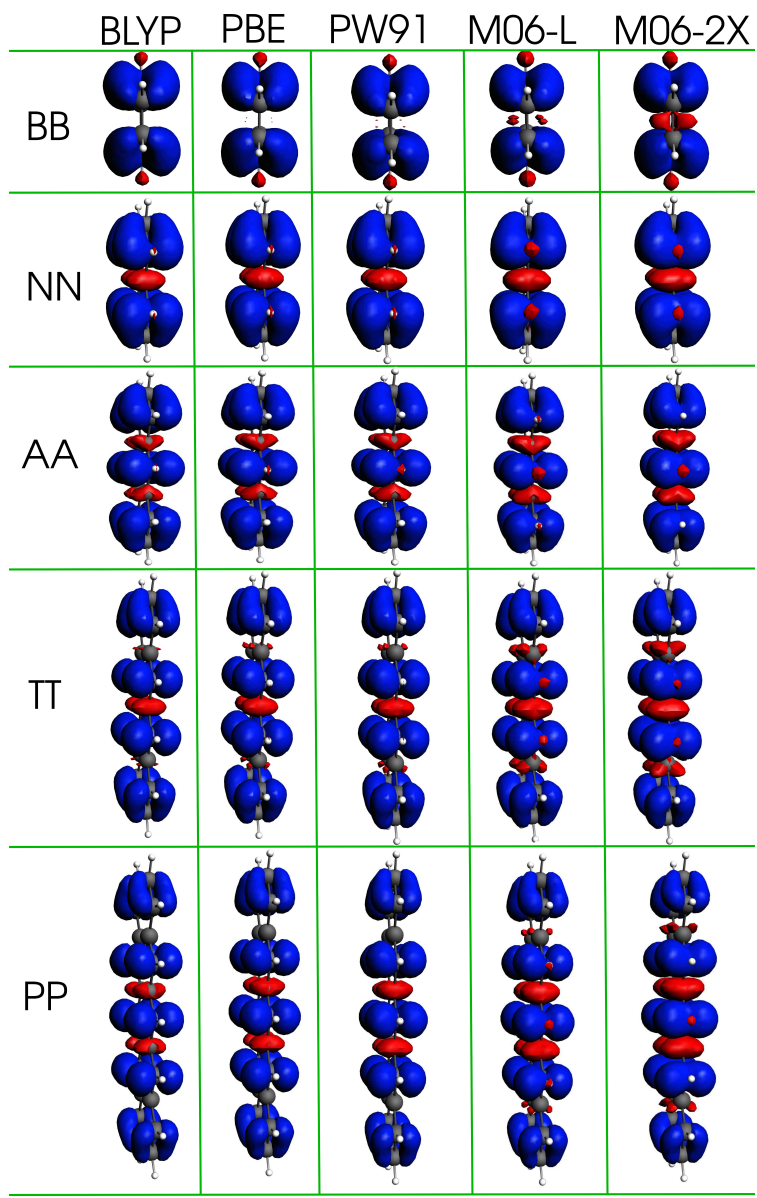}
\end{center}
\caption{Progressive polarization of the spin densities calculated for benzene (BB), naphthalene (NN), anthracene (AA), tetracene (TT) and pentacene (PP). Each column corresponds to a different XC functional. Benzene's spin density is the most sensitive.}
\label{sdpic}
\end{figure}
A different picture is presented for BB, in Figures \ref{muefun1} and \ref{mruefun1}, and for AC and BB, AA and TT in Figure S3. In these cases, even though the large majority of the XC/NAKE/basis set combinations are in good to excellent agreement with the benchmark results, some functional/basis set combination did not compare quantitatively to the benchmark. Looking at Figures \ref{muefun1} and \ref{mruefun1} it is obvious that BB is the more problematic of the systems for BLYP. All other XC functionals perform well for this dimer. BLYP is surprisingly inaccurate at long and short ranges, with either small or large basis set. Although for this functional only, many possibilities were tried in order to improve the performance, for instance, additional freeze-and-thaw cycles were run (3 cycles are the default), a finer integration grid and more accurate density fitting were also employed with no achieved improvements. This can be related to the known difficulties of semilocal functionals to model open shell systems \cite{Pople1995,cremer2002}. For benzene and its derivatives, we notice that different functionals have an effect on the spin density polarization of the radical cation susbsystem, as exemplified by the spin-density plots in Figure \ref{sdpic}. The BLYP functional produces the least spin polarized systems. A similar effect was reported for DNA nucleobase dimers \cite{solo2012} where the spin density polarization was more pronounced for MP2 than for GGA XC functionals. 
In addition, we found the SCF to be slowly converging for BB, especially when BLYP is employed, as in the BB radical cation there is a degeneracy that is difficult to lift. Our claim is that this singular behavior of BLYP is related to its inability to match higher level of theory models for the radical cation subsystem, in turn undermining its ability to produce quantitatively correct couplings for BB.

We do not provide here an explanation for the failure of metahybrids and metaGGAs for the AC system at long ranges. Use of these functionals in conjunction with FDE is so far untested and more investigations are needed to shed light on their behavior.

\begin{figure}[htp]
\begin{center}
\includegraphics[width=0.8\textwidth]{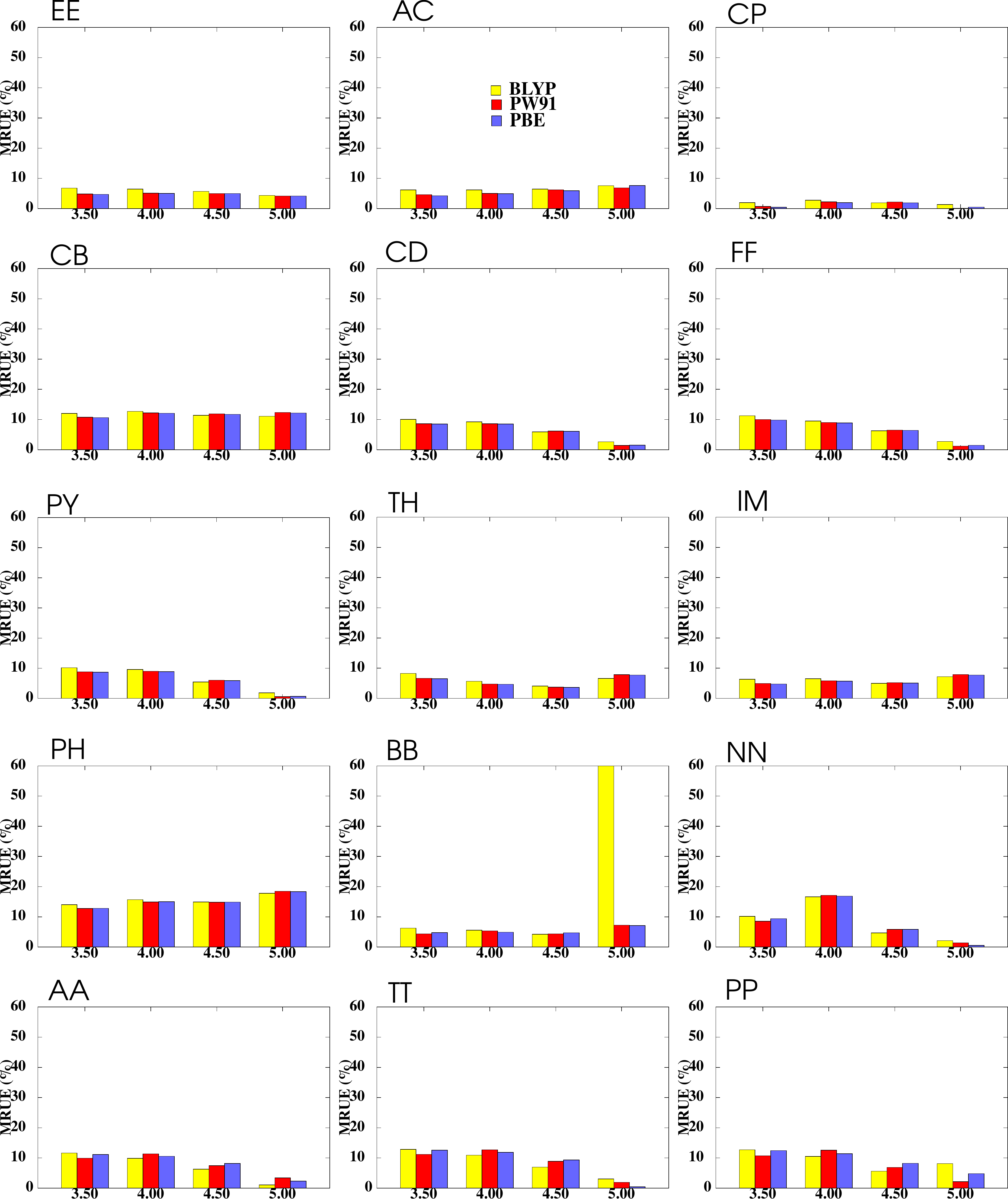} 
\end{center}
\caption{MRUE in \% for the performance of GGA XC functionals. TZP and PW91k are employed.}
\label{mruefun1}
\end{figure}

\begin{figure}[htp]
\begin{center}
\includegraphics[width=0.8\textwidth]{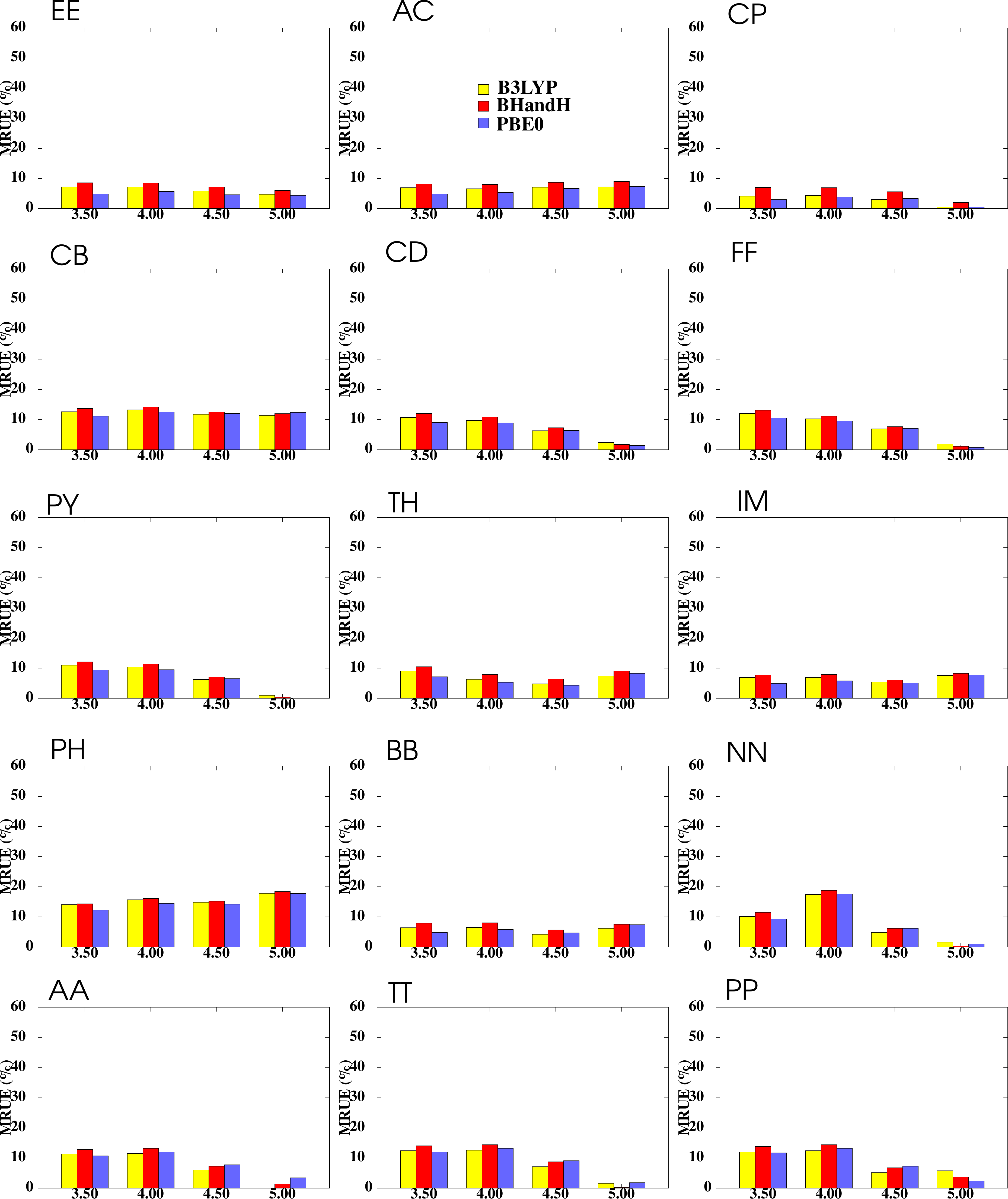} 
\end{center}
\caption{MRUE in \% for the performance of hybrid XC functionals. TZP and PW91k are employed.}
\label{mruefun2}
\end{figure}

\begin{figure}[htp]
\begin{center}
\includegraphics[width=0.8\textwidth]{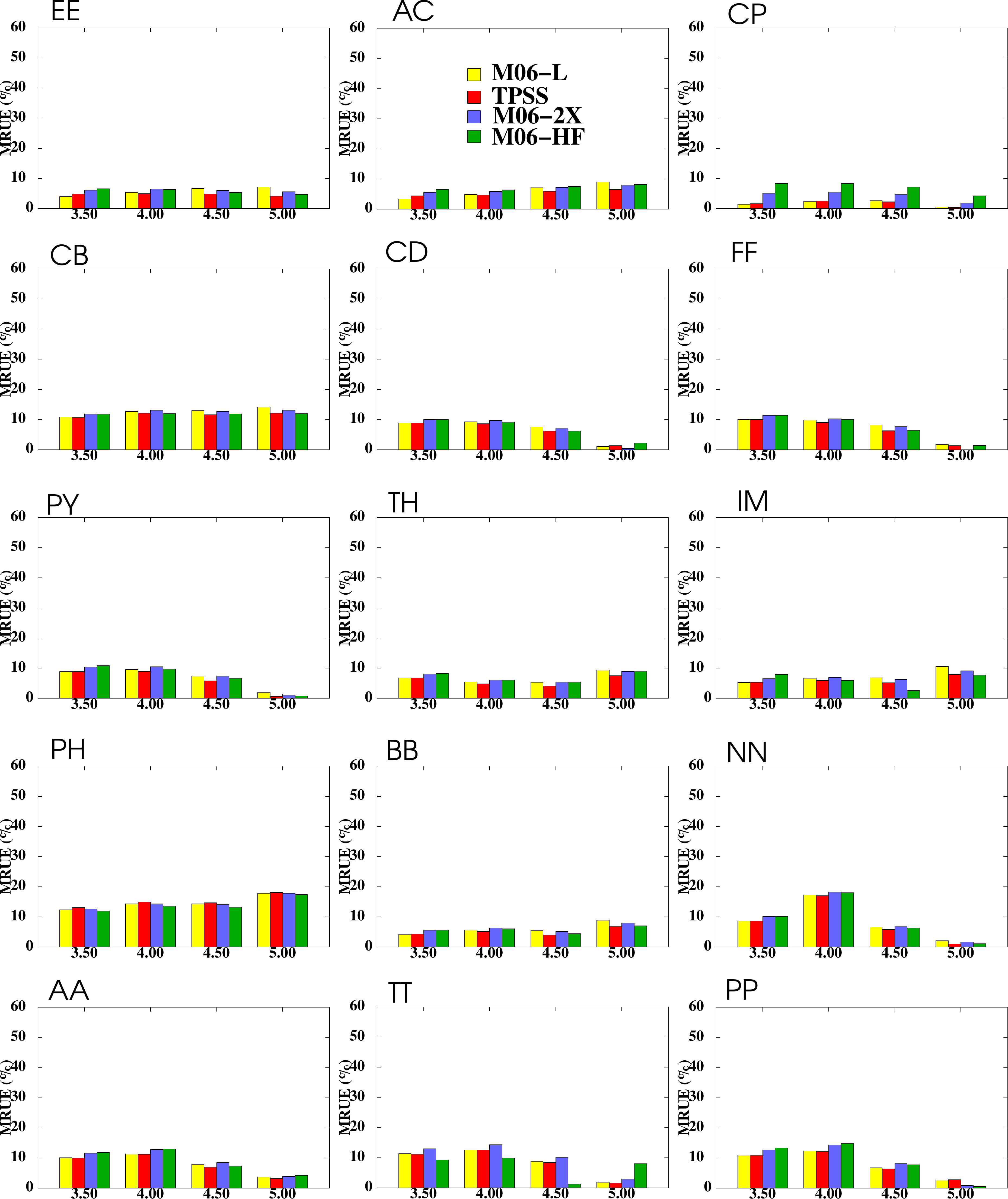} 
\end{center}
\caption{MRUE in \% for the performance of metaGGA and metahybrid XC functionals. TZP and PW91k are employed. }
\label{mruefun3}
\end{figure}
In conclusion, GGA functionals are generally a good choice, with the unique exemption of BLYP for the benzene dimer. Table \ref{bests} collects the best methods (i.e., the more stable for different system sizes) and PBE is reported as being the most accurate and transferable functional. The results for GGAs are in good agreement with the benchmark values, and in some cases they showed to be superior to non-local and MetaGGA functionals. Besides GGAs, B3LYP stands out as another valuable choice.
\begin{table}[htp]
\begin{center}
\begin{tabular}{crrr}
 \toprule

Set & MUE(meV) & MRUE(\%) & MAX(meV)\\
 \hline

PBE/PW91k/TZP    & 15.3 & 7.1 & 49.6\\

PW91/PW91k/TZP   & 18.0 & 8.3 & 48.4\\

B3LYP/PW91k/TZP  & 18.5 & 8.0 & 27.6\\
M06-2X/PW91k/TZP & 29.1 & 14.1 & 90.0\\
 \bottomrule
\end{tabular}
\end{center}
\caption{Mean stistical values for the best XC-functional choices.}
\label{bests}
\end{table}
\subsection{Electronic coupling dependence at different rotational angles}
\label{rota}
In this section we focus on the sensitivity of the FDE method when the dimers are placed at different geometry configurations. The calculations were performed on ethylene and thiophene dimers whose geometries were borrowed from Ref.\cite{Kubas2014}. The ethylene dimer was tested by rotating one ethylene molecule around the center of mass at 5 \AA\ intersubsystem separation. On the other hand, thiophene was classified in three different type of rotations. First in a sandwich configuration of the dimer, one thiophene molecule rotates around the center of mass, alternatively, the two thiophene molecules were rotated around the rotation axes that passes through the S atom, finally, one thiophene rotates randomly around the center of mass. Intermolecular distances of 5 \AA, 6.75 \AA \ and 4 \AA\ were considered in order to keep a minimal distance between the hydrogen atoms of the dimer (for more details about these rotations on ethylene and thiophene see Figure 4 in Ref.\ \citenum{Kubas2014}).
 
\begin{figure}[t]
\begin{center}
\includegraphics[width=0.8\textwidth]{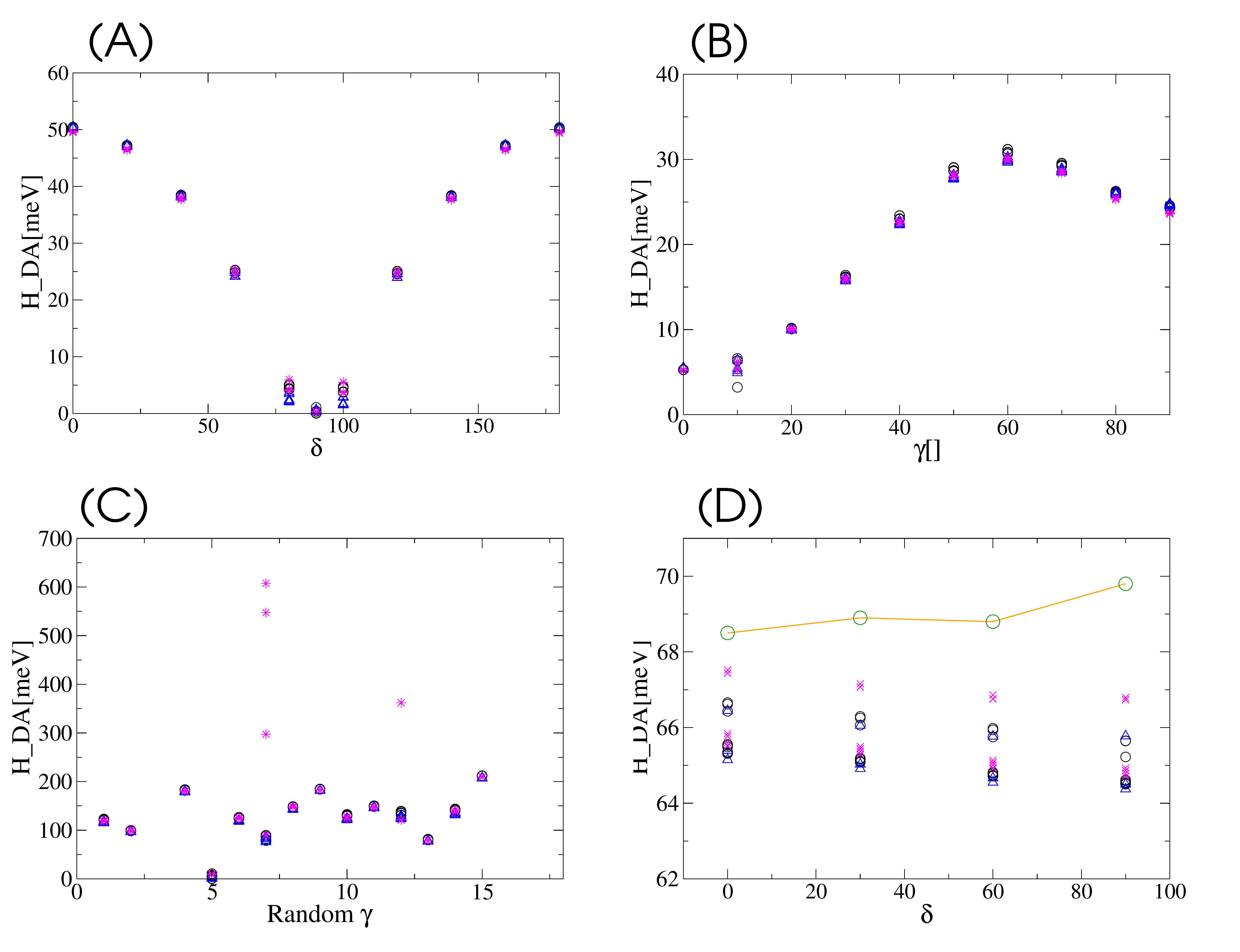} 
\end{center}
\caption{Behavior of electronic coupling in thiophene dimer at (a) sandwich configuration rotations, (b) simultaneous disrotation of the dimer and (c) random rotations, and (d) ethylene dimer. In each case the set of PBE functional (black circle), B3LYP (blue triangle) and M06-2X (magenta star) were used in conjunction with PW91k functional and TZP basis set. The orange line in (d) correspond to MRCI+Q.}
\label{allrot}
\end{figure}

In Figure \ref{allrot} we report the performance of FDE-ET varying the XC functionals, as this was the one category that featured the largest deviations for the HAB11 previously studied. In the supplementary information, the analysis w.r.t. the basis set and NAKE functionals is also reported. Regarding the EE system (see Figure \ref{allrot}d), all functionals show appreciable deviations at $90{^\circ}$ rotation angle. This is because the two double bonds are perpendicularly placed to each other and a nodal structure arises such that the overlap between the diabats is small. Because of this, numerical inaccuracies creep in the inversion of the transition overlap matrix to compute the transition density. Such a problem was detected already by us in a past study \cite{ramos2014} and for which we proposed a solution based on the Penrose inversion of the transition overlap matrix. In this study, however, we purposely did not report values obtained adjusting this threshold. However, upon adjusting the threshold to a lower value, also the couplings at $90{^\circ}$ compared quantitatively with the benchmark. In related methods, alternatives to the Penrose inversion have been proposed \cite{Evan2013,Tsaune1994289,Colle1987,Glush2010}. 

Generally, all functionals perform satisfactorily. The PBE and PW91 GGA functionals are in overall good agreement. However, there is a dependence on the basis set: the larger the basis set, the more accurate the coupling. In Figure S4, the basis set dependence of the electronic coupling with respect to the rotation angle is shown, although TZP and TZ2P perform well, the QZ4P basis set seems to yield the best results. Possibly because the distance between the ethylene monomers is larger than 4.5 \AA\ and thus in the long-range where we saw before the QZ4P performs best.

The second test case is comprised of three different kinds of rotations of the thiophene dimer, see Figures \ref{allrot}a--\ref{allrot}c. In these three cases, the couplings are strongly dependent on the angle. The dependence is clearly due to the orbital overlap of the subsystem HOMOs, whenever there is a nodal structure (cancellation of different phases of the orbitals) the coupling will become negligible. By comparison of Figure 5 and Table IX on Ref.\ \citenum{Kubas2014} with the present results, the FDE-ET method ranks at the same level of CDFT with coupling values a bit lower than the reference. Figure \ref{allrot}a, shows that for the case considered, the coupling seem to be relatively independent form the chosen functional, or basis set (see Figure S4), and all functionals yield couplings within a few meV from the benchmark values.

\section{Conclusions}
\label{conclu}
The most important finding in this work resides in the fact that GGA functionals coupled with a medium sized basis set and the PW91k NAKE functional allows the FDE-ET method to yield reliable electronic couplings as tested against high-level correlated wavefunction methods applied to an array of donor-acceptor dyads. We find the PBE functional to be the most transferable functional in each case considered having a MAX error lower than 30 meV and an overall MRUE of a little over 7\%. This constitutes a success for the FDE-ET method. 

We analyze the performance of 10 XC functionals, ranging from GGAs to the Minnesota meta GGA functionals, and also hybrid functionals with Hartree-Fock exchange ranging from 10-30\%, and metahybrid functionals with HF exchange in the 50-100\% range. We extract from the statistics that the XC functionals are determinant in the performance of the electronic coupling. Conversely, the NAKE functionals statistically do not play an important role (e.g., the couplings are relatively insensitive to their choice). In addition, our analysis of the basis set dependence shows that the QZ4P basis set (the largest set considered) is the most problematic as it often undermines the FDE convergence at short intermonomer separations -- a problem already well documented in the FDE literature \cite{jaco2007,Fradelos2011,Fradelos2011b}. 

Overall, we show that by varying the three parameters considered in this study: XC functionals, NAKE functionals and basis sets, diabatic states are correctly generated with FDE-ET. In addition to the quality of the diabats, we provide convincing computational evidence that the FDE-ET method produces couplings which satisfactorily correlate with the benchmark data. In conclusion, the FDE-ET is found to be a powerful tool for modeling CT (specifically hole transfer) reactions.

\section{Acknowledgements}
This work was funded by an grant from the National Science Foundation, grant No.\ CBET-1438493.

\bibliography{literatur}

\end{document}